\documentclass[10pt,conference]{IEEEtran}
\IEEEoverridecommandlockouts

\usepackage{amsmath,amssymb,amsfonts}
\usepackage{algorithmic}
\usepackage{graphicx}
\usepackage{textcomp}
\usepackage{xcolor}
\usepackage{algorithmic}
\usepackage[ruled,vlined,linesnumbered]{algorithm2e}
\newcommand{\makeunderscoreletter}{\catcode`\_=11}

\usepackage{syntax}
\usepackage[shortlabels]{enumitem}

\usepackage{cite}
\usepackage{balance}
\usepackage{url}

\def \toolname {AndroEvolve}

\definecolor{red}{HTML}{9B0000}
\definecolor{lightred}{HTML}{FF5131}
\definecolor{green}{HTML}{006400}
\definecolor{lightgreen}{HTML}{9CFF57}
\definecolor{purple}{HTML}{7200CA}
\definecolor{verylightgrey}{HTML}{F1F1F1}
\definecolor{diffstart}{named}{blue}
\definecolor{diffincl}{named}{green}
\definecolor{diffrem}{named}{red}
\usepackage{listings}
  \lstdefinelanguage{diff}{
	basicstyle=\ttfamily\extrabold\scriptsize,
	morecomment=[f][\color{diffstart}]{@},
	morecomment=[f][\color{diffincl}]{+},
	morecomment=[f][\color{diffrem}]{-},
        keepspaces=true,
	identifierstyle=\color{black},
  }
\newboolean{showcomments}
\setboolean{showcomments}{true}
\ifthenelse{\boolean{showcomments}}
{ \newcommand{\mynote}[2]{\textcolor{orange}{
			\fbox{\bfseries\sffamily\scriptsize#1}
			{\small$\blacktriangleright$\textsf{\emph{#2}}$\blacktriangleleft$}}}}
{ \newcommand{\mynote}[2]{}}

\ifthenelse{\boolean{showcomments}}
{ \newcommand{\dnote}[2]{\textcolor{red}{
			\fbox{\bfseries\sffamily\scriptsize#1}
			{\small$\blacktriangleright$\textsf{\emph{#2}}$\blacktriangleleft$}}}}
{ \newcommand{\dnote}[2]{}}

\ifthenelse{\boolean{showcomments}}
{ \newcommand{\hnote}[2]{\textcolor{blue}{
			\fbox{\bfseries\sffamily\scriptsize#1}
			{\small$\blacktriangleright$\textsf{\emph{#2}}$\blacktriangleleft$}}}}
{ \newcommand{\hnote}[2]{}}

\def\BibTeX{{\rm B\kern-.05em{\sc i\kern-.025em b}\kern-.08em
    T\kern-.1667em\lower.7ex\hbox{E}\kern-.125emX}}
\begin{document}
\makeunderscoreletter
\title{AndroEvolve: Automated Update for Android Deprecated-API Usages\\
}



\author{
\IEEEauthorblockN{
Stefanus A. Haryono\IEEEauthorrefmark{1},
Ferdian Thung\IEEEauthorrefmark{1},
David Lo\IEEEauthorrefmark{1},
Lingxiao Jiang\IEEEauthorrefmark{1},
Julia Lawall\IEEEauthorrefmark{3},
Hong Jin Kang\IEEEauthorrefmark{1},\\
Lucas Serrano\IEEEauthorrefmark{2}, and
Gilles Muller\IEEEauthorrefmark{3}
}

\IEEEauthorblockA{\IEEEauthorrefmark{1}School of Information Systems, Singapore Management University, Singapore\\
\{stefanusah,ferdianthung,davidlo,hjkang.2018,lxjiang\}@smu.edu.sg}
\IEEEauthorblockA{\IEEEauthorrefmark{2}Sorbonne University/Inria/LIP6, France\\
Lucas.Serrano@lip6.fr}
\IEEEauthorblockA{\IEEEauthorrefmark{3}Inria, France\\
\{Gilles.Muller,Julia.Lawall\}@inria.fr}
}


\def \toolname {AndroEvolve}

\maketitle
\thispagestyle{plain}
\pagestyle{plain}

\begin{abstract}
The Android operating system (OS) is often updated, where each new version may involve API deprecation. Usages of deprecated APIs in Android apps need to be updated to ensure the apps' compatibility with the old and new versions of the Android OS. 
In this work, we propose \toolname{}, an automated tool to update usages of deprecated Android APIs, that addresses the limitations of the state-of-the-art tool, CocciEvolve. \toolname{} utilizes data flow analysis to solve the problem of out-of-method-boundary variables, and variable denormalization to remove the temporary variables introduced by CocciEvolve. We evaluated the accuracy of \toolname{} using a dataset of 360 target files and 20 deprecated Android APIs, where \toolname{} is able to produce 319 correct updates, compared to CocciEvolve which only produces 249 correct updates. We also evaluated the readability of \toolname{}'s update results using a manual and an automatic evaluation. Both evaluations demonstrated that the code produced by \toolname{} has higher readability than CocciEvolve's.
A video demonstration of \toolname{} is available at \url{https://youtu.be/siU0tuMITXI}.
\end{abstract}

\section{Introduction}
Android is one of the most widely used operating systems (OS) in the recent years due to the popularity of smartphones. It is frequently updated to add new features and fix bugs. Following each OS update, changes and modifications in the Android APIs are common, which often results in deprecation of the older version of an API and makes the deprecated APIs unusable for the newer version of the OS. This problem, named Android fragmentation~\cite{han2012understanding, wei2016taming}, is a common occurrence.

Due to the above issue, updating usages of deprecated Android APIs should be prioritized. Several studies have promoted automated approaches to updating deprecated Android API usages. 
A4~\cite{lamothe_a4} automatically assists Android API migrations by learning API migration patterns from the differences between the deprecated and updated code examples. AppEvolve~\cite{fazzini2019automated} uses both before- and after- update code examples to automatically learn the update. However, a replication study by Thung et al.~\cite{thung_towardsgenerating} found several weaknesses in AppEvolve, where the target file must be structured the same way as the code examples.

Following Thung et al.'s findings, Haryono et al.~\cite{coccievolve} presented CocciEvolve, an automated deprecated Android API usage update tool built on top of Coccinelle4J~\cite{kang2019automating}. 
CocciEvolve offers several improvements from AppEvolve: (1) a readable update script expressed in the Semantic Patch Language (SmPL)~\cite{lawall2018coccinelle}; (2) only requiring a single after-update example; and (3) the ability to update multiple invocations of a deprecated API within a single file. 
Code example used in CocciEvolve must be in the form of an {\tt if} statement, containing the updated API in the {\tt then} statement and the old API in the {\tt else} statement, or vice versa. An example of such a code example is given in Fig.~\ref{fig:example_update_code}, where the deprecated API {\tt getCurrentMinute} is shown at line 5 and the updated API {\tt getMinute} at line 3.

\vspace{-3.0 mm}

\begin{figure}[h]
	\centering
	\scriptsize{
\begin{lstlisting}[xleftmargin=5.0ex,language=java,numbers=left,sensitive=true,columns=flexible,basicstyle=\ttfamily]
if (android.os.Build.VERSION.SDK_INT >= 
        android.os.Build.VERSION_CODES.M) {
    minutes = picker.getMinute();
} else {
    minutes = picker.getCurrentMinute();
}
\end{lstlisting}
\vspace{-3.0 mm}
        \setlength{\belowcaptionskip}{7pt}
		\caption{An example of after-update code for {\tt getCurrentMinute()} API}\label{fig:example_update_code}
	}
\end{figure}

While CocciEvolve significantly improves on AppEvolve, it has several shortcomings. CocciEvolve is only able to resolve (i.e., locate and identify the definition for) the value of a variable within the same method, meaning that if an expression used in the updated API invocation is defined outside of the currently processed method, CocciEvolve will not be able to resolve its value. 
Furthermore, during the update process, CocciEvolve introduces temporary variables that refer to other existing variables in the target file.
These temporary variables are not deleted in the update result, making the update result less readable.

In this paper, we present \toolname{}, an automated Android deprecated API usage update tool that addresses the limitations of CocciEvolve. This paper is a tool demo paper that supplements our full paper~\cite{haryono2020androevolve}. \toolname{} addresses the limitations through two features: data flow analysis and variable denormalization. Data flow analysis is used to resolve the values used as the updated API invocation arguments, including values of all out-of-method variables. Using the data flow analysis, \toolname{} automatically locates the definitions of those variables, and uses their values to replace the variables used in the API invocation arguments. Variable name denormalization refers to the process that reverts the normalization done by CocciEvolve that introduces the temporary variables in the update result. This process reverts the temporary variables back to their original variable names within the code, removing all unnecessary changes and improving the overall readability of the update result. 

We evaluated \toolname{} on a dataset of 360 target files containing 20 deprecated Android APIs. 
Based on this evaluation, AndroEvolve produces 316 correct updates while CocciEvolve only produces 249 correct updates, showing that AndroEvolve outperforms CocciEvolve by 26.90\%.
We also measure the readability of the update results of both tools using manual and automatic evaluation methods.
For the manual evaluation, we asked the opinions of two experienced Android engineers. For the automated evaluation, we use a popular readability scoring tool~\cite{readabilitymodel}. We found that \toolname{}'s update results are 83\% and 50\% more readable compared to CocciEvolve's according to the manual and automated readability evaluation, respectively.

\section{Design and Implementation}
\subsection{\toolname{} Architecture}
\begin{figure}[t]
	\centering  
	\includegraphics[width=0.85\linewidth]{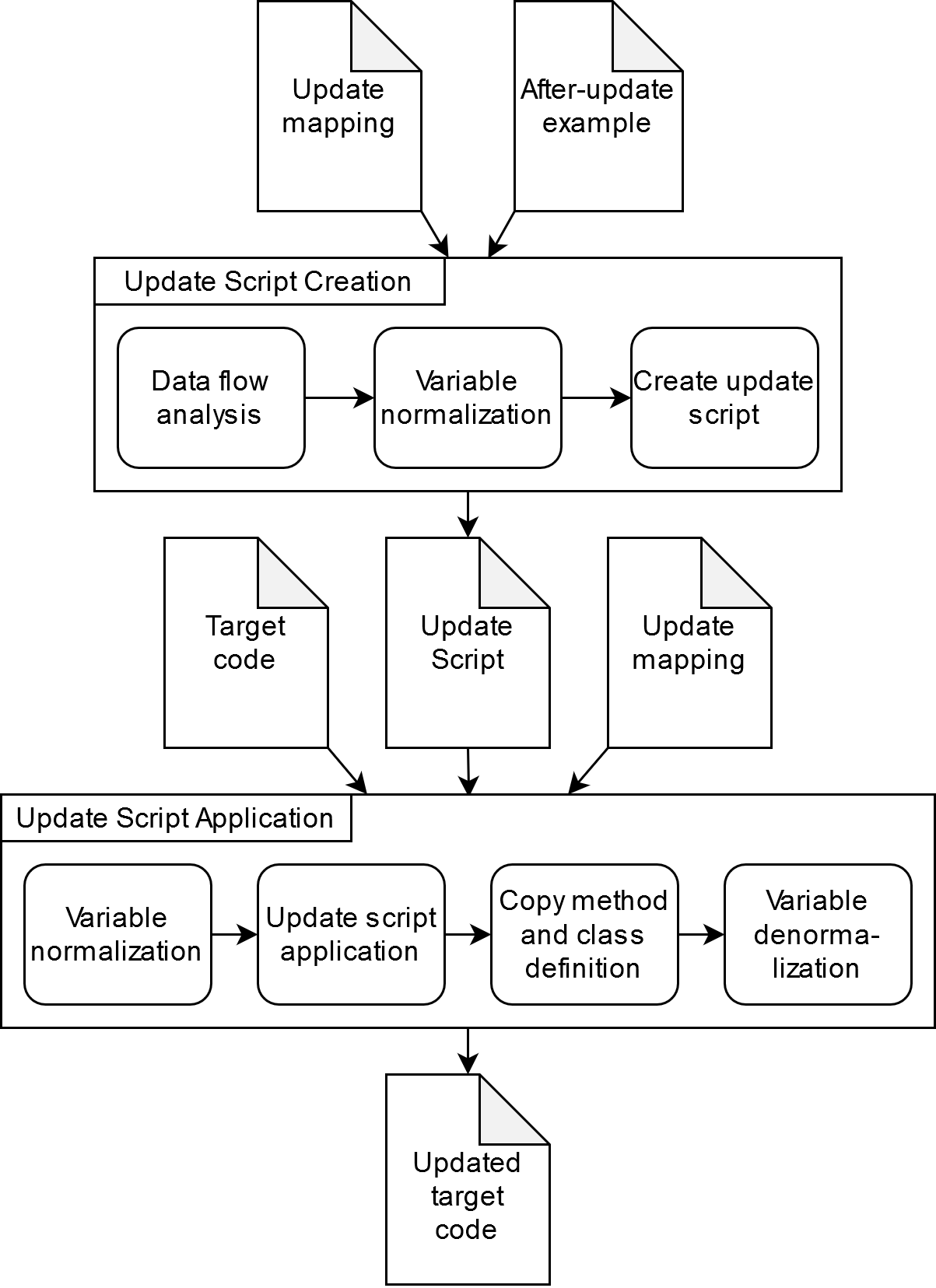}
	\caption{Summary of \toolname{} workflow}
	\label{fig:androevolve_summary}
	\vspace{-3.0mm}
\end{figure}

The overall architecture of \toolname{} is shown in Fig.~\ref{fig:androevolve_summary}.
The workflow of \toolname{} is comprised of {\em update-script creation} followed by {\em update-script application}.

\textbf{Update-script creation} takes as input the API update mapping and an after-update example of the API. The API update mapping consists of the API signatures of the deprecated API and its corresponding updated API. This mapping is used to identify the deprecated and the updated API from the code example and target code.
Next, data flow analysis is employed to resolve the values of out-of-method variables in the update example that are used as the arguments of the updated API invocation but are not used in the deprecated API invocation. Variable normalization is then applied to the part of the code containing the deprecated and updated API invocations, introducing temporary variables to name all complex subexpressions to facilitate the update. The goal is to minimize the syntactic differences between the update example and the target code to be updated. The update script, expressed using the Semantic Patch Language (SmPL), is created from the normalized update example.

\textbf{Update-script application} takes as input the target code to be updated, the update script from the update-script creation process, and the API update mapping. The update-script application is also comprised of several steps. First, variable normalization is applied into the target code.
Then, the update script is applied to the normalized target code using Coccinelle4J~\cite{kang2019automating}, resulting in the updated code. Following the update script application, we copy method and class definitions that are being used in the updated API arguments to the updated code. Finally, variable denormalization is applied to this updated code, replacing the temporary variables with their original expressions.

The following subsections further describe these components. The complete details of each \toolname{} component and process are available in our full paper~\cite{haryono2020androevolve}.

\subsection{Data Flow Analysis}
\toolname{} utilizes a data flow analysis (DFA) to resolve the values of expressions used as arguments in an API method invocation. 
The data flow analysis is able to gather and predict a set of possible values of an object/variable at any given point inside the code, thus it is able to predict and resolve the correct replacement values for any expressions used in the updated API invocation arguments.

We build the DFA using the symbol resolver from Java Symbol Solver that is available in Javaparser.\footnote{https://javaparser.org/} The DFA performs a bottom-up search from the variables or expressions used as the API invocation arguments and expands the search scope until it finds the value or methods referred to by the expressions. Values and method invocations that are found by this DFA are then used to replace the original expressions used as API invocation arguments to ensure that the API arguments are in the form of either a literal expression, a static class member, a method invocation, or an object creation.

If the found values are in the form of method invocations or object creations that are not available in the target code, it is necessary to copy the method or class definition from the after-update example to the target file to ensure that it is accessible in the updated code. For this purpose, if the resolved API invocation arguments are in the form of method invocations or object creations, \toolname{} will extract their definitions and copy them to the target file during the update. \toolname{} only copies the methods and classes that are related to the resolved API arguments.

\subsection{Variable Denormalization}

Variable normalization introduces temporary variables for each API invocation argument which clutter the update result and reduce its readability. 
Using variable denormalization, \toolname{} aims to remove these temporary variables and replace them with their values or referenced variables. For each temporary variable, \toolname{} locates its definition and replaces it with the resolved values or expressions. Then, \toolname{} deletes the declarations and definitions of the temporary variables as they are no longer needed.


\section{Evaluation}
\subsection{Dataset and Experiment Setup}
Our dataset is comprised of three components: after-update examples used for the update-script creation, one-to-one API mappings from the deprecated APIs to the replacement APIs, and the target files to be updated.
We utilize the after-update examples that were used in the evaluation of AppEvolve~\cite{fazzini2019automated}.
For the target files to be updated, we extend the target file dataset from CocciEvolve~\cite{coccievolve}. We randomly selected public GitHub projects obtained using AUSearch~\cite{asyrofi2020ausearch}, a tool to search GitHub repositories for API usages. 
We collected 360 target files containing 20 deprecated Android APIs for our target files dataset.

We evaluated the performance of \toolname{} by comparing its update accuracy (i.e., the percentage of correct updates) against CocciEvolve's. 
We also compare the readability of the update results of the two tools, using an automated and a manual approach. In the automated approach, we utilized a state-of-the-art automated readability measuring tool from Scalabrino et al.~\cite{readabilitymodel} which gives a score of 0.0-1.0 where higher score indicates better readability. 
In the manual approach, we asked two experienced Android developers to score 60 successful updated code fragments, with 30 updated code fragments from each of CocciEvolve and \toolname{}. We choose 30 as the sample size to represent the variation in the syntax of the updated code. The developers were not told which tool was used to produce each updated code fragment. For each updated code fragment, the developers were asked to give score in the Likert scale of 1-5 for the readability and the naturalness of the code. A higher score indicates higher readability and higher confidence that the code resembles code produced by a human.

\subsection{Experiment Results}
We manually check the correctness of the updated code produced by both \toolname{} and CocciEvolve. On the dataset, \toolname{} and CocciEvolve are able to provide correct updates for 316 and 249 target files, respectively. Our analysis shows that data flow analysis helps to improve the number of correct updates produced by \toolname{}. It works particularly well in cases where the after-update example uses out-of-method variables. 

In the readability evaluation, using the automatic approach, we measure the readability scores of the updated code and calculate the average for each API. We found that \toolname{}'s updated code achieves higher readability than CocciEvolve's for all of the evaluated deprecated API migrations. This finding is strengthened by our manual readability measurement where the average readability score given by the developers for \toolname{} achieves a score of 4.817, while CocciEvolve only achieves a score of 2.633. Similarly, the code naturalness measure for \toolname{} achieves an average score 4.917, while CocciEvolve achieves an average score of 2.433. 
These evaluations highlight that \toolname{} offers a significant improvement in code readability as compared to CocciEvolve.

\section{Usage Example}
\toolname{} can be used via a command line interface (CLI). 
Two main functionalities are provided:
\begin{enumerate}[nosep,leftmargin=*]
    \item \textbf{Update Patch Creation}
    Update patch creation takes as input the API update mapping and the after-update example. \toolname{}'s CLI takes the command line argument {\tt --generate-patch} to specify the update patch creation, {\tt --input} to indicate the path for the update example file, and {\tt --output} to indicate the path for the generated patch file. The format of this command is as follows:

\begin{scriptsize}
\begin{lstlisting}[language=diff,sensitive=true,columns=flexible,basicstyle=\ttfamily]
(*\textbf{java -jar AndroEvolve.jar --generate-patch}*)
    <deprecated_api_signature> <updated_api_signature> 
    (*\textbf{--input}*) <update_example_path>
    (*\textbf{--output}*) <output_path>
\end{lstlisting}
\end{scriptsize}
    
    \item \textbf{Update Patch Application}
    Update patch application takes as input the API update mapping, the target file to be updated, and the previously created update patch file. \toolname{}'s CLI takes the command line argument {\tt --apply-patch} to specify the update patch application process, {\tt --input} to indicate the path to the target file, {\tt --patch} to indicate the path to the update patch file, and {\tt --output} to indicate the path for the produced updated target file. The format of this command is as follows:
    
\begin{scriptsize}
\begin{lstlisting}[language=diff,sensitive=true,columns=flexible,basicstyle=\ttfamily]
(*\textbf{java -jar AndroEvolve.jar --apply-patch}*)
    <deprecated_api_signature> <updated_api_signature>
    (*\textbf{--input}*) <target_filepath> (*\textbf{--patch}*) <patch_filepath>
    (*\textbf{--output}*) <output_path>
    
\end{lstlisting}
\vspace{-5.0mm}
\end{scriptsize}
    
\end{enumerate}

\begin{figure}[h]
	\centering
	\scriptsize{
\begin{lstlisting}[xleftmargin=5.0ex,language=java,numbers=left,sensitive=true,columns=flexible,basicstyle=\ttfamily]
private static final int DURATION = 50;
private static final int AMPLITUDE = 175;
public static void itemActivated(Context context) {
    long milliseconds = DURATION;
    if (android.os.Build.VERSION.SDK_INT >= 
            android.os.Build.VERSION_CODES.O) {
    	int amplitude = AMPLITUDE;
    	VibrationEffect effect = VibrationEffect.
    	        createOneShot(milliseconds, amplitude);
        (*\textbf{vibrator.vibrate(effect);}*)
    } else {
        (*\textbf{vibrator.vibrate(milliseconds);}*)
    }
}
\end{lstlisting}
\vspace{-3.0mm}
        \setlength{\belowcaptionskip}{7pt}
		\caption{After-update example for {\tt android.os.Vibrator\#vibrate} {\tt (long)} deprecated API}\label{fig:after-update-example}
	}
\end{figure}

As an example of the patch creation and update application process provided by \toolname{}, consider the {\tt android.os.Vibrator\#vibrate(long)} deprecated API. This deprecated API needs to be updated into {\tt android.os.Vibrator\#vibrate(android.os.Vi-} {\tt brationEffect)}, meaning that a change to the API parameter is necessary. An after-update example is provided in Fig.~\ref{fig:after-update-example}. 
Line 10 contains the updated API usage, while line 12 contain the deprecated API usage. Using this update example, we use the following command to run the update patch creation process:

\begin{scriptsize}
\begin{lstlisting}[language=diff,sensitive=true,columns=flexible,basicstyle=\ttfamily]
(*\textbf{java -jar AndroEvolve.jar --generate-patch}*) 
    android.os.Vibrator#vibrate(long)
    android.os.Vibrator#vibrate(android.os.VibrationEffect)
    (*\textbf{--input}*) example.java
    (*\textbf{--output}*) vibrate_update.cocci
\end{lstlisting}
\end{scriptsize}

In line 10, we can see that the updated API invocation uses the variable {\tt effect} as its argument. The data flow analysis conducted a bottom-up search in the example for this variable, finding its definition at lines 8-9. The search is continued as the {\tt effect} variable definition uses other variables, i.e, {\tt milliseconds} and {\tt amplitude}. This search is done recursively until all variables related to the API invocation arguments are resolved. Following this data flow analysis, variable normalization is conducted, and the update patch is created. The created update patch is shown in Fig.~\ref{fig:update_patch}. 

\vspace{-3.0 mm}

\begin{figure}[h]
	\centering
	\scriptsize{
\begin{lstlisting}[xleftmargin=5.0ex,language=diff,numbers=left,sensitive=true,columns=flexible,basicstyle=\ttfamily]
@bottomupper_classname@
expression exp0, exp1;
identifier iden0, classIden;
@@
...
+ if (android.os.Build.VERSION.SDK_INT >= 
+       android.os.Build.VERSION_CODES.O) {
+ VibrationEffect newParameterVariable0 = 
+       VibrationEffect.createOneShot(50, 175);
+ classIden.vibrate(newParameterVariable0);
+ } else {
classIden.vibrate(iden0);
+ }
\end{lstlisting}
\vspace{-3.0 mm}
        \setlength{\belowcaptionskip}{7pt}
		\caption{Update patch based on the after-update example in Fig.~\ref{fig:after-update-example} for the {\tt android.os.Vibrator\#vibrate(long)} deprecated API}\label{fig:update_patch}
	}
\end{figure}


Using this created update patch, we can apply the update to a target file containing the deprecated API usage.
The following command runs the update application process:

\begin{scriptsize}
\begin{lstlisting}[language=diff,sensitive=true,columns=flexible,basicstyle=\ttfamily]
(*\textbf{java -jar AndroEvolve.jar --apply-patch}*)
    android.os.Vibrator#vibrate(long)
    android.os.Vibrator#vibrate(android.os.VibrationEffect)
    (*\textbf{--input}*) main.java (*\textbf{--patch}*) vibrate_update.cocci
    (*\textbf{--output}*) main_updated.java
\end{lstlisting}
\end{scriptsize}

The target file is first normalized to remove syntactic differences from the update patch. Then, the update patch is applied, resulting in the updated code. Finally, variable denormalization is applied to remove the temporary variables that are introduced in the normalization. Fig.~\ref{fig:update_result} highlights the code differences resulting from the application of this update. An {\tt if} statement that checks the Android version is added at line 3. Lines 5-7 contain the updated {\tt android.os.Vibrator\#vibrate(android.os.Vi-} {\tt brationEffect)} API usage, along with the new parameter required for its invocation.

\balance
\section{Conclusion}
We have proposed \toolname{} as the next step in the automated update of usages of Android deprecated APIs. \toolname{} provides a significant improvement compared to the state-of-the-art tool, CocciEvolve, through the addition of data flow analysis to handle out-of-method variables, and variable denormalization to increase the readability of the update result by removing temporary variables previously introduced by CocciEvolve. We evaluated the accuracy and readability of \toolname{} and demonstrated its higher accuracy and readability compared to CocciEvolve. \toolname{} is available at \url{https://github.com/soarsmu/AndroEvolve}

\noindent{\bf Acknowledgement.} This research is supported by the Singapore NRF (award number: NRF2016-NRF-ANR003) and the ANR ITrans project.

\vspace{-3.0 mm}
\begin{figure}[h]
	\centering
	\scriptsize{
\begin{lstlisting}[xleftmargin=5.0ex,language=diff,numbers=left,sensitive=true,columns=flexible,basicstyle=\ttfamily]
public void Once(long milliseconds) {
    if (MyVibrator.hasVibrator()) {
+       if (Build.VERSION.SDK_INT >= 
+           Build.VERSION_CODES.O) {
+           MyVibrator.vibrate(VibrationEffect.
+               createOneShot(500, VibrationEffect.
+               DEFAULT_AMPLITUDE));
+       } else {
            MyVibrator.vibrate(milliseconds);
+       }
    }
}
\end{lstlisting}
\vspace{-3.0 mm}
        \setlength{\belowcaptionskip}{7pt}
		\caption{Update result for a target file using the update patch in Fig.~\ref{fig:update_patch}}\label{fig:update_result}
	}
\end{figure}
\vspace{-2.0mm}




\bibliography{references}
\bibliographystyle{IEEEtran}

\end{document}